\documentclass[11pt]{article}
\usepackage[dvipsnames,svgnames,x11names,hyperref]{xcolor}
\usepackage{amsmath}
\usepackage{amsthm}
\usepackage[T1]{fontenc}

\usepackage{fullpage}
\usepackage{thmtools}
\usepackage[colorlinks]{hyperref}
\hypersetup{
  linkcolor=[rgb]{0,0,0.4},
  citecolor=[rgb]{0, 0.4, 0},
  urlcolor=[rgb]{0.6, 0, 0}
}
\usepackage{amssymb,amsfonts,amsmath,amsthm,amscd,dsfont,mathrsfs,bbm}
\usepackage{todonotes}
\usepackage{complexity}
\usepackage{tikz}
\usetikzlibrary{decorations.pathreplacing,backgrounds}
\usepackage{multirow}

\usepackage{amsthm}
\usepackage{thmtools,thm-restate}

\numberwithin{equation}{section}
\declaretheoremstyle[bodyfont=\it,qed=\qedsymbol]{noproofstyle}

\declaretheorem[numberlike=equation]{observation}

\declaretheorem[name=Observation,numbered=no]{observation*}

\declaretheorem[numberlike=equation]{fact}

\declaretheorem[numberlike=equation]{theorem}

\declaretheorem[name=Theorem,numbered=no]{theorem*}

\declaretheorem[numberlike=equation]{lemma}
\declaretheorem[name=Lemma,numbered=no]{lemma*}

\declaretheorem[numberlike=equation]{corollary}
\declaretheorem[name=Corollary,numbered=no]{corollary*}

\declaretheorem[name=Proposition,numbered=no]{proposition*}

\declaretheorem[numberlike=equation]{claim}
\declaretheorem[name=Claim,numbered=no]{claim*}

\declaretheorem[name=Conjecture,numbered=no]{conjecture*}

\declaretheorem[name=Question,numbered=no]{question*}

\declaretheoremstyle[bodyfont=\it,qed=$\lozenge$]{defstyle} 

\declaretheorem[numberlike=equation,style=defstyle]{definition}
\declaretheorem[unnumbered,name=Definition,style=defstyle]{definition*}

\declaretheorem[unnumbered,name=Example,style=defstyle]{example*}

\declaretheorem[unnumbered,name=Notation=defstyle]{notation*}

\declaretheorem[unnumbered,name=Construction,style=defstyle]{construction*}

\declaretheorem[unnumbered,name=Remark,style=defstyle]{remark*}

\usepackage{nth}
\usepackage{intcalc}
\usepackage{etoolbox}
\usepackage{xstring}

\usepackage{ifpdf}
\ifpdf
\else
\usepackage[quadpoints=false]{hypdvips}
\fi

\newcommand{\ECCCurl}[2]{http://eccc.hpi-web.de/report/\ifnumcomp{#1}{>}{93}{19}{20}#1/#2/}

\newcommand{\shortECCC}[2]{\texttt{\href{http://eccc.hpi-web.de/report/\ifnumcomp{#1}{>}{93}{19}{20}#1/#2/}{eccc:TR#1-#2}}}

\newcommand{\parseECCC}[1]{% Takes a string of the form TRxx/xxx or
\StrSubstitute{#1}{TR}{}[\tmpstring]%
\IfSubStr{\tmpstring}{/}{ %assuming string is of the form TRxx/xxx
\StrBefore{\tmpstring}{/}[\ecccyear]%
\StrBehind{\tmpstring}{/}[\ecccreport]%
}{% assuming string is of the form TRxx-xxx
\StrBefore{\tmpstring}{-}[\ecccyear]%
\StrBehind{\tmpstring}{-}[\ecccreport]%
}%
\shortECCC{\ecccyear}{\ecccreport}}

\newcommand{\F}{\mathbb{F}}
\newcommand{\N}{\mathbb{N}}

\newcommand{\va}{\mathbf{a}}
\newcommand{\vb}{\mathbf{b}}
\newcommand{\vc}{\mathbf{c}}
\newcommand{\ve}{\mathbf{e}}

\newcommand{\vx}{\mathbf{x}}

\newcommand{\vz}{\mathbf{z}}

\newcommand{\vzero}{\mathbf{0}}

\newcommand{\diag}{\mathsf{diag}}

\newcommand{\Det}{\mathsf{Det}}

\newcommand{\rank}{\mathsf{rank}}

\DeclareMathOperator{\Sing}{Sing}
\DeclareMathOperator{\Img}{Im}

\DeclareMathOperator{\dc}{\mathsf{dc}}
\DeclareMathOperator{\rdc}{\mathsf{rdc}}

\DeclareMathOperator{\Hom}{\mathsf{Hom}}

\DeclareMathOperator{\abp}{\mathsf{abp}} %abp size
\DeclareMathOperator{\habp}{\mathsf{habp}} %homogeneous abp size
\DeclareMathOperator{\abpw}{\mathsf{abp_w}} % homogeneous abp width
\DeclareMathOperator{\habpw}{\mathsf{habp_w}} % homogeneous abp width

\newcommand{\set}[1]{\left\{ #1 \right\}}

\newclass{\VBP}{VBP}

\title{Determinants vs.\ Algebraic Branching Programs}
\author{Abhranil Chatterjee\thanks{Indian Statistical Institute, Kolkata, India. Email: \texttt{abhneil@gmail.com}. Research supported by the DST-INSPIRE Faculty Fellowship.}
\and
Mrinal Kumar\thanks{Tata Institute of Fundamental Research, Mumbai, India. Email: \texttt{mrinal@tifr.res.in}.  Research supported by the Department of Atomic Energy, Government of India, under project 12-R{\&}D-TFR-5.01-0500.}
\and
 Ben Lee Volk\thanks{Efi Arazi School of Computer Science, Reichman University, Israel. Email: \texttt{benleevolk@gmail.com}. }}

\date{}

\begin{document}

\maketitle

\abstract{
We show that for every homogeneous polynomial of degree $d$, if it has determinantal complexity at most $s$, then it can be computed by a homogeneous algebraic branching program (ABP) of size at most $O(d^5s)$.
 Moreover, we show that for \textit{most} homogeneous polynomials, the width of the resulting homogeneous ABP is just $s-1$ and the size is at most $O(ds)$.
  
Thus, for constant degree homogeneous polynomials, their determinantal complexity and  ABP complexity are within a constant factor of each other and hence, a super-linear lower bound for ABPs for any constant degree polynomial implies a super-linear lower bound on determinantal complexity; this relates two open problems of great interest in algebraic complexity. As of now, super-linear lower bounds for ABPs are known only for polynomials of growing degree \cite{K19, CKSV22}, and for determinantal complexity the best lower bounds are larger than the number of variables only by a constant factor \cite{KV22}.

While determinantal complexity and ABP complexity are classically known to be polynomially equivalent~\cite{MV97}, the standard transformation from the former to the latter incurs a polynomial blow up in size in the process, and thus, it was unclear if a super-linear lower bound for ABPs implies a super-linear lower bound on determinantal complexity. In particular, a size preserving transformation from determinantal complexity to ABPs does not appear to have been known prior to this work, even for constant degree polynomials. 
}

\section{Introduction}
\label{sec:intro}

\subsection{Super-linear Lower Bounds in Complexity Theory}
\label{subsec:super-linear}

Perhaps the principal embarrassment of complexity theory at the present time is its failure to provide techniques for proving non-trivial lower bounds on the complexity of some of the commonest combinatorial and arithmetic problems. To add further to the embarrassment, the previous sentence is a direct quote from Valiant's 1975 paper \cite{Valiant75}, and yet it is true today as it was the day it was written, nearly 50 years ago: we are still unable to prove, for example, a super-linear circuit lower bound for a problem in $\NP$. 

Proving such lower bounds for boolean circuits is a major open problem in complexity theory (even for circuits of depth $O(\log n)$), but such an analogous result is known in the model of \emph{algebraic circuits}, that compute multivariate polynomials using arithmetic operations. Baur and Strassen \cite{Strassen73, BS83} proved that computing the degree-$d$ power sum polynomial $\sum_{i=1}^n x_i^d$, for instance, requires circuits of size $\Omega(n \log d)$, which is super-linear in $n$ assuming $d=\omega(1)$.

As evident from the statement of the result (and even more so from the proof technique), this result crucially relies on the fact that the polynomial being computed is of high degree. It remains an interesting open problem to prove super-linear lower bounds for algebraic circuits computing constant degree polynomials (see \cite{raz-elusive} for a discussion on the importance of this problem). In fact, even the case of computing linear transformations has attracted significant attention (see, e.g., \cite{Valiant77, ShoupS1997, Lokam09}).

\subsection{Algebraic Branching Programs}
\label{subsec:intro-abps}

Circuits are the most powerful model of algebraic computation, and therefore one may consider first solving this challenge in easier settings. A \emph{formula} is a circuit whose underlying graph is a tree. Kalorkoti \cite{Kal85} developed a technique for proving super-linear lower bounds for algebraic formulas (based upon Nechiporuk's method which applies to Boolean formulas \cite{Nec66}). This technique can be used to prove lower bounds as large as $\Omega(n^2  / \log n)$ for mutilinear polynomials of degree $n$ (as discussed in \cite{CKSV22}, in the model of algebraic formulas it is natural to consider multilinear polynomials since allowing polynomials of large individual degree results in some trivial lower bounds). This lower bound is slightly improved in \cite{CKSV22}, using a different technique, to $\Omega(n^2)$. It is fairly straightforward to apply Kalorkoti's method to constant degree polynomials as well and obtain super-linear lower bounds.

Lying between formulas and circuits is the model of \emph{algebraic branching programs} (or ABPs, which are defined formally in \autoref{sec:prelim}). The best lower bound known for ABPs is $\Omega(nd)$ \cite{CKSV22}, which is again only super-linear when the degree $d$ is super-constant. Interestingly, in the analogous boolean model of branching programs (even parity or non-deterministic branching programs, which are arguably a better suiting analog of algebraic branching programs), Nechiporuk's method can be used to show super-linear lower bounds (see, e.g., \cite{KW93}). In the algebraic setting, however, the ability to label a single edge of the program by an arbitrary linear function of the variables seems like a challenge to this technique. Therefore, it is still an open problem to prove super-linear lower bounds for constant-degree polynomials, not only for circuits but even for algebraic branching programs.

The family of multivariate polynomials that can be computed by ABPs of polynomial size form the complexity class $\VBP$. One notable member of this class is the determinant polynomial, $\Det(X) = \sum_{\sigma \in S_n} \prod_{i=1}^n x_{i,\sigma(i)}$. It is, in fact, a rather distinguished member of this class: perhaps the most natural way to define ``reductions'' between polynomials is to consider linear projections of the variables, and under this class of reductions the determinant is a \emph{complete} polynomial for $\VBP$, namely, every polynomial in this class can be efficiently reduced to the determinant (see, e.g., \cite{S15} for a proof of this fact). Arguably, the fact that determinants are ubiquitous in mathematics can be attributed to this completeness result, as any polynomial with a small branching program (and in particular, any polynomial with a small formula) is just a determinant in disguise.

\subsection{Determinantal Complexity}
\label{subsec:intro-dc}

The discussion in the previous paragraph implies that one can equivalently define $\VBP$ using \emph{determinantal representations}. A determinantal representation of size $s$ for a polynomial $f \in \F[x_1, \ldots, x_n]$ is an $s \times s$ matrix $M$ whose entries are linear functions\footnote{Throughout this text when we use the term ``linear functions'' to include affine functions as well. When we insist that the constant term is zero, we make it explicit by referring to \emph{homogeneous} linear functions.} in $x_1, \ldots, x_n$ such that $\Det(M) = s$. The completeness result mentioned above in particular implies that the determinantal complexity of every polynomial is finite, and also motivates studying the \emph{determinantal complexity} of polynomials as a natural complexity measure of its own.

For a polynomial $f$, let $\dc(f)$ denote its determinantal complexity, that is, the minimal $s$ such that $f$ has a determinantal representation of size $s$. By the discussion above, proving that the determinantal complexity of a sequence of polynomials $\set{f_n}$ is super-polynomial (in the number of variables $n$) immediately implies that this sequence is not in $\VBP$. The sad reality, however, is that we don't know a single \emph{explicit} sequence of polynomials whose determinantal complexity is super-linear (it is easy to show that this quantity is exponential for a ``generic'' polynomial, or even a random polynomial with 0/1 coefficients). The best lower bound, as a function of the number of variables $n$, is roughly $1.5n$, proved by Kumar and Volk \cite{KV22}. Also worth mentioning is the lower bound of Mignon and Ressayre \cite{MR04}, who proved that the determinantal complexity of the $n \times n$ permanent over fields of characteristic 0 is at least $n^2/2$. This result was extended to characteristics different than 2 by Cai, Chen and Li \cite{CCL10} (over characteristic 2, the permanent and determinant coincide). Note however that the $n \times n$ permanent is a polynomial in $n^2$ variables, so this lower bound isn't super-linear in the number of variables, and in fact, it seems challenging to extend their technique to prove a lower bound which is larger than the number of variables. We refer to the introduction of \cite{KV22} for further discussion on this subject.

The super-linear lower bounds for algebraic branching programs (or even circuits) don't imply super-linear lower bounds on determinantal complexity, as the known reductions to the determinant incur a polynomial blow-up in the size parameter: that is, a polynomial computed by a size-$s$ ABP also has a determinantal representation of size $\poly(s) \times \poly(s)$, but if the best lower bound we can give on $s$ is slightly super-linear, the lower bound we get on the size of the determinantal representation isn't super-linear in $n$.

\subsection{Our Contributions}
\label{sec:results}

We relate here the two problems mentioned above, of proving lower bounds for constant degree polynomials and proving lower bounds for determinantal complexity, by reducing one to the other. Our main result is that proving a super-linear lower bound on the ABP complexity of a homogeneous constant-degree polynomial would imply a super-linear lower bound on its determinantal complexity. The other direction of this reduction is already known: a polynomial that has a size-$s$ ABP can be written as the determinant of an $s \times s$ matrix whose entries are linear functions \cite{S15}, so a super-linear lower bound on $\dc(f)$ also implies a lower bound on its ABP complexity.

We prove this reduction by constructing an ABP of size $\poly(d) \cdot s$ for any homogeneous polynomial that has a size-$s$ determinantal representation.

\begin{theorem}[Informal]
\label{thm:intro:general}
Let $f$ be a homogeneous polynomial that has a size-$s$ determinantal representation. Than $f$ has an ABP of size $\poly(d) \cdot s$.
\end{theorem}

The formal statement of \autoref{thm:intro:general} appears as \autoref{thm:gen-det-abp} in \autoref{sec:general}.

We stress again that while the fact that any polynomial with small determinantal representation has a small ABP isn't new, the known reductions from determinants to ABPs (which are simply constructions of algebraic branching programs for the determinant, e.g., \cite{B84, MV97}) all result in ABPs of size $s^{1+c}$ for some $c>0$. Thus, \autoref{thm:intro:general} gives a more efficient reduction when $d$ is small, and in particular we have the following corollary.

\begin{corollary}
\label{cor:general}
Let $f = \set{f_n}$ be a family of homogeneous polynomials of constant degree. Then a super-linear lower bound on the ABP complexity of $f$ implies a super-linear lower bound on the determinantal complexity of $f$.
\end{corollary}

\autoref{thm:intro:general} applies for every constant degree polynomial. It turns out, however, that for ``most'' polynomials we can construct a reduction which is simultaneously simpler and more efficient. We first explain what do we mean by ``most'' polynomials. The \emph{singular locus} of a homogeneous polynomial $f$ is the variety defined by the common zeros of its first order partial derivatives. This variety is defined by $n$ polynomials and thus for a ``generic'' homogeneous polynomial $f$, one expects this variety to be zero-dimensional. There are, of course, notable exceptions. For the $n \times n$ determinant polynomial, for example (that has $n^2$ variables), this variety has dimension exactly $n^2 - 4$ (see \cite{vzG87}).

Having a singular locus whose dimension is strictly less than $n-4$ imposes powerful structure on the determinantal representations which we are able to use in order to prove:

\begin{theorem}[Informal]
\label{thm:intro:generic}
Let $f$ be a homogenous polynomial such that its singular locus has dimension less than $n-4$, and $f$ has a size-$s$ determinantal representation. Than $f$ has an ABP of width $s-1$ and size at most $ds$.
\end{theorem}

The formal statement of \autoref{thm:intro:general} appears as \autoref{cor:generic-det-abp} (of \autoref{thm:reg-det-abp}) in \autoref{sec:regular}.

The structure that we exploit to prove \autoref{thm:intro:generic} is the fact that for polynomials whose singular locus has dimension less than $n-4$, it holds that
 the constant part of any determinantal representation must be of rank $s-1$ (this fact was discovered by \cite{vzG87} and is also used in the determinantal complexity lower bounds of \cite{ABV17} and \cite{KV22}: see \autoref{obs:regular-reps} in \autoref{sec:prelim}). This motivated Landsberg and Ressayre \cite{LR17} to define the notion of \emph{regular determinantal complexity}. A determinantal representation is said to be regular if its constant part has rank $s-1$, and the regular determinantal complexity of a polynomial $f$ is the size of its smallest regular determinantal representation. Another motivation for this definition comes from the fact that the natural reductions from formulas and ABPs to determinantal representations give regular determinantal representations. \autoref{thm:reg-det-abp} shows that the notions of regular determinantal complexity and ABP complexity are essentially equivalent, since the relation between the size of the regular determinantal representation and the width of the ABP is particularly sharp.

We stress again that ``almost all'' polynomials satisfy the assumption of \autoref{thm:intro:generic}. In particular, it seems conceivable that most proof techniques that would imply a lower bound for ABPs will be applicable to at least one polynomial with a small dimensional singular locus, so that we can apply \autoref{thm:intro:generic} to obtain lower bounds on its determinantal complexity.

A final remark is in order regarding the homogeneity assumptions in \autoref{thm:intro:general} and \autoref{thm:intro:generic}. We don't see it as a major hurdle towards proving lower bounds using our reduction. Most polynomials studied in algebraic complexity (such as the determinant, permanent, elementary symmetric polynomials, power sum polynomials, iterative matrix multiplication, and so on) are homogeneous to begin with, and we are not aware of a single technique for proving lower bounds that uses non-homogeneity in a crucial way, so it seems conceivable again that most proof techniques for lower bounds would apply to homogeneous polynomials just as well.

Nevertheless, it is still an assumption, and one may try to eliminate it, if only for purely aesthetic reasons. One natural way to go about it is to try to handle each homogeneous component of $f$ separately, apply our transformation to obtain an ABP, and then assemble the ABPs computing each homogeneous component to an ABP computing the sum.

Such an approach indeed works in other contexts in algebraic complexity theory, where non-homogeneity is rarely an issue when studying strong models of computation such as ABPs and circuits. These models can be efficiently homogenized: namely, given a possibly non-homogeneous circuit or ABP computing a polynomial $f$, one can find, for each homogeneous component of $f$, a circuit (or ABP) that computes it, whose size is bigger than the original size only by a multiplicative factor which depends polynomially on the degree $d$ (even further, one can find a \emph{single} circuit or ABP with multiple outputs, that \emph{simultaneously} computes all the homogeneous components, with similar size guarantees). One can then handle each homogeneous component of $f$ separately. Thus, when one considers super-polynomial lower bounds this is never an issue, and for the question of proving super-linear lower bound this isn't a problem if $d$ is a constant.

Curiously enough, however, we don't know if the same can be done for determinantal representation. While it is true that if $f$ has a size $s$ determinantal representation then each of its homogeneous component has a determinantal representation of size $\poly(s,d)$ (where $d = \deg(f)$), the proof for this fact involves first converting the determinantal representation to an ABP using, for example, the reductions of \cite{MV97, B84}, homogenizing the ABP, and converting the ABP back to a determinantal representation. This results in a size blow-up which is super-linear in $s$ (due to the first step of the transformation), which makes it unsuitable for us.

A similar issue arises when one considers determinantal complexity of sums of polynomials, which leads to the question of whether $\dc(f+g) \le \dc(f) + \dc(g)$ (or perhaps the inequality holds up to a constant factor). It is easy to see that $\dc(f \cdot g) \le \dc(f) + \dc(g)$ (just take a block matrix), and using the conversion to ABPs it's also easy to conclude that $\dc(f+g) = \poly(\dc(f), \dc(g))$, but as most natural models of computation have the stronger subadditivity property, it would be interesting to prove it (or disprove it) for determinantal complexity.

\section{Preliminaries}
\label{sec:prelim}

In this section we give definitions of some of the notions we use later, and state some basic results.

We start by defining the singular locus of a polynomial.

\begin{definition}\label{def:sing-locus}
Let $f \in \F[\vx]$ be a polynomial. The hypersurface defined by $f$, $V(f)$, is the set of points $\va$ such that $f(\va) = 0$. The singular locus of $f$, $\Sing(f)$ is the variety defined by
\[
\Sing(f) = \left\{ \va \in V(f) : \frac{\partial f}{\partial x_i}(\va) = 0, 1\leq i \leq n \right\}.\qedhere
\]
\end{definition}

We briefly remark that some previous related papers (such as \cite{CKSV22, KV22}) defined the singular locus as simply the set of common zeros of the first order partial derivatives of $f$, without requiring that they are also zeros of $f$. In this context this is a minor distinction that has no significance on the results of \cite{CKSV22, KV22} or the results of this paper. However in \autoref{sec:r-regular} we consider a generalization of \autoref{def:sing-locus} to higher order derivatives and thus it is slightly more elegant to use the definition above.

\begin{fact}[\cite{vzG87}]\label{fact:vzgathen}
Let $\F$ be an algebraically closed field and let $\Det_m$ be the $m \times m$ determinant polynomial. Then $\Sing(\Det_m) \subseteq \F^{m\times m}$ is precisely the set of matrices of rank at
most $m - 2$, and $\dim \Sing(\Det_m) = m^2 - 4$.
\end{fact}

We also require the following elementary and well known identity.

\begin{lemma}\label{lem:schur-complement}
Let $M \in \F^{m\times m}$ be a matrix over a field $\F$ and let $A, B, C, D$ be submatrices of $M$ of dimension $k \times k, k \times (m-k), (m-k)\times k$ and $(m-k) \times (m-k)$ respectively, such that
\[ M = \begin{pmatrix}
   A & B\\\
C & D
 \end{pmatrix} \, .
 \]
If the matrix $D$ is invertible, then 
 \[
 \Det (M)  = \Det (A - BD^{-1}C ) \cdot \Det (D) \, .
 \]
\end{lemma}

\subsection{Determinantal Complexity}

We now define the determinantal complexity of a polynomial $f$.

\begin{definition}\label{def:det-complexity}
Let $f \in \F[x_1, \ldots, x_n]$ be a polynomial of degree $d$. A \emph{determinantal representation} of $f$ of size $s$ is an $s \times s$ matrix $M$, whose entries are linear functions in $x_1, \ldots, x_n$, such that $\Det(M) = f$. We denote by $\dc(f)$ the minimal integer $s$ such that $f$ has a determinantal representation of size $s$.

A determinantal representation $M$ of $f$ is said to be \emph{regular} if the constant part $M_0$ of $M$ (i.e., $M(\vzero)$) is of rank $s-1$. We denote by $\rdc(f)$ the minimal integer $s$ such that $f$ has a regular determinantal representation of size $s$.
\end{definition}

The fact that the determinantal complexity is finite for every polynomial $f$ was established by Valiant \cite{Val79}. As explained in \cite{LR17}, the same proof establishes the fact that the regular determinantal complexity is also finite (as several of the proofs of Valiant's theorem construct regular determinantal representations).

In fact, the following lemma of von zur Gathen \cite{vzG87} shows that for ``most'' homogeneous polynomials one may consider regular determinantal representations without loss of generality.
 We refer to \cite{vzG87, ABV17, KV22} for proofs of this fact, and \cite{LR17} for some related discussion.

\begin{lemma}\label{lemma:vzg}
Let $f \in \F[\vx]$ be a polynomial, and let $M : \F^n \to \F^{s\times s}$ be a polynomial map
such that $f(\vx) = \Det_s(M(\vx))$. Suppose further that $\dim(\Sing(f)) < n - 4$. Then $\Img(M) \cap
\Sing(\Det_s) = \emptyset$. Furthermore, all matrices in $\Img(M)$ have rank at least $s - 1$.
\end{lemma}
An easy consequence of this lemma is the following observation. 
\begin{observation}
\label{obs:regular-reps}
Let $f \in \F[\vx]$ be a polynomial whose constant term is zero. Further assume that $\dim(\Sing(f)) < n - 4$. Then any determinantal representation of $f$ must be regular.
\end{observation}

\begin{proof}
Let $f$ has a determinantal representation $M$ of size $s$. As $\dim(\Sing(f)) < n - 4$, using \autoref{lemma:vzg} we know that all matrices in $\Img(M)$ have rank at least $s - 1$. In particular, $M(\vzero)$ is of rank $\geq s - 1$. $M(\vzero)$ can not be of full rank as $f$ has a zero constant term, so $f(\vzero) = \Det(M(\vzero)) = 0$. Therefore, the rank of $M(\vzero)$ is exactly $s - 1$ and $M$ is a regular determinantal representation of $f$.
\end{proof}

\subsection{ABP complexity}

We now define the ABP complexity of a polynomial $f$. As our work deals with the fine notions of complexity (rather than separating polynomial size from super polynomial size), we take care to account for the various subtleties concerning the definition.

\begin{definition}\label{def:ABP}
Let $f \in \F[x_1, \ldots, x_n]$ be a polynomial of degree $d$. We say $f$ has an algebraic branching program (ABP) of \emph{width} $w$ and \emph{size} $s$ if
\[
f = \vb^T M_1 \cdot M_2 \cdots M_k \vc,
\]
where $\vb \in \F^{w_0}$ and $\vc \in \F^{w_k}$ are vectors whose entries are linear functions in $x_1, \ldots, x_n$, for every $i \in [k]$, $M_i \in \F^{w_{i-1} \times w_i}$  are matrices whose entries are linear functions in $x_1, \ldots, x_n$, and the following properties hold:
\begin{enumerate}
\item $w_i \le w$ for all $0 \le i \le [k]$
\item $\sum_{i=0}^k w_i \le s$.
\end{enumerate}
We say that the ABP is \emph{homogeneous} if all functions appearing in $\vb, \vc$ and the $M_i$'s are homogeneous.

We denote by $\abp(f)$ the minimal $s$ such that $f$ has a size-$s$ ABP (of any width), and by $\abpw(f)$ the minimal $w$ such that $f$ has a width-$w$ ABP (of any size). We similarly use $\habp(f)$ and $\habpw(f)$ for the homogeneous variants of these notions.
\end{definition}

\autoref{def:ABP} is an algebraic definition. Equivalently, one may define algebraic branching programs in a graph-theoretic equivalent way, as a labeled, layered and directed acyclic graph, with a source and a sink, in which the matrices above correspond to the adjacency matrices between one layer to the next. The graph theoretic definition is more convenient when one considers operations on ABPs such as taking their sum or their product, homogenizing them or composing them. Note that our measure of ``size'' counts vertices and not edges.

Note that for a homogeneous polynomial $f$, $\abp(f) \le \habp(f) \le (d+1) \cdot \abp(f)$, where the first inequality is trivial and the second follows by a standard homogenization argument, and a similar inequality holds for ABP width. More formally, we have the following lemma.  

\begin{lemma}[Partial homogenization of ABPs]\label{lem:abp-partial-homogenization}
Let $A$ be an ABP of size $s$ and width $w$ computing a polynomial $f$ of degree $\Delta$. Then, for every $d \in \{0, 1, \ldots, \Delta\}$,  there exists a homogeneous ABP $\tilde{A}_d$ of size at most $s(d+1)$ and width at most $w(d+1)$ that computes the homogeneous component of degree $d$ of $f$. 
\end{lemma}

Finally, note that in a homogeneous ABP we must have exactly $d+1$ layers, and $\habp(f) \le (d+1) \cdot \habpw(f)$.

We also use the following result of Mahajan and Vinay \cite{MV97} who showed that determinants have small ABPs. 
\begin{theorem}[\cite{MV97}]\label{thm:ABP-for-det}
For every field $\F$ and all $n \in \N$, the polynomial $\Det_n$ can be computed by an ABP of width $n^2$ and $(n+1)$ layers, and thus total size $O(n^3)$.
\end{theorem}

\section{Algebraic Branching Programs from Regular Determinantal Representation}
\label{sec:regular}

In this section we prove our results for regular determinantal representations. Recall again that by \autoref{obs:regular-reps}, for ``most'' polynomials, one may consider regular representations without loss of generality, as all of their determinantal representation are regular.

The following theorem states that the regular determinantal complexity of a homogeneous polynomial $f$ is an upper bound on its homogeneous ABP width.

\begin{theorem}
\label{thm:reg-det-abp}
Let $f \in \F[x_1, \ldots, x_n]$ be a homogeneous polynomial of degree $d\ge2$. Suppose $\rdc(f) = s$. Then $\habpw(f) \le s-1$ (and in particular, $\abp(f) \le \habp(f) = O(ds) $).
\end{theorem}

As an immediately corollary of \autoref{obs:regular-reps} and \autoref{thm:reg-det-abp}, we obtain:

\begin{corollary}
\label{cor:generic-det-abp}
Let $f \in \F[x_1, \ldots, x_n]$ be a homogeneous polynomial of degree $d\ge2$ such that $\dim(\Sing(f)) < n-4$. Then $\habpw(f) \le s-1$ (and in particular, $\abp(f) \le \habp(f) = O(ds) $).
\end{corollary}

\begin{proof}[Proof of \autoref{thm:reg-det-abp}]
Let $M(\vx)$ be a regular determinantal representation of $f$ of size $s$, that is, $\Det(M)=f$. Write $M = M'(\vx) + M_0$ where $M_0$ is the constant part of $M$, which is of rank $s-1$, and $M'$ is a matrix whose entries are homogeneous linear functions. As in \cite{vzG87, ABV17, KV22}, by applying elementary row and column operations we may assume that $M_0 = \diag(0,1,\ldots,1)$. Thus, we can write $M$ in blocks as
\[
M = \begin{bmatrix}
a & \vb^T \\
\vc & I - D
\end{bmatrix}
\]
where $a(\vx)$ is a homogeneous linear polynomial, $\vb, \vc \in \F^{s-1}$ are vectors of homogeneous linear polynomials, and $D \in \F^{(s-1) \times (s-1)}$ is a matrix of homogeneous linear polynomials. We now claim that $f = -\vb^T (D^{d-2}) \vc$, which implies the statement of the theorem.

To see this, note that $I-D$ is an invertible matrix over $\F(\vx)$ (indeed, its determinant is a polynomial whose constant term is 1, so it is non-zero), and therefore by \autoref{lem:schur-complement},
\begin{equation}\label{eq:determinant-block}
f = \Det(M) = \Det(I-D) \cdot \Det(a - \vb^T (I-D)^{-1} \vc) = \Det(I-D) \cdot (a - \vb^T (I-D)^{-1} \vc).
\end{equation}
The last equality follows from the fact that $a - \vb^T (I-D)^{-1} \vc$ is a $1 \times 1$ matrix.

We write $\Det(I-D) = 1 + R$ where $R$ is a polynomial whose constant term is zero. We also note that we can expand $(I-D)^{-1}$ as a power series $(I-D)^{-1} = I + D + D^2 + D^3 + \cdots $ over the ring of formal power series $\F[[\vx]]$, and thus, \[(a - \vb^T (I-D)^{-1} \vc) = a - \vb^T I \vc - \sum_{i \geq 1}\vb^T D^i \vc  \] (this equality holds in the ring $\F[[\vx]]$).

In particular, the homogeneous component of degree $0$ of the power series above is zero, the degree one homogeneous component is $a$, degree two homogeneous component is $ \vb^T \vc $ and for every $i > 2$, the degree $i$ homogeneous component equals $\vb^T D^{i-2} \vc$. 

Plugging this into \eqref{eq:determinant-block}, we get
\begin{equation}
\label{eq:det-power-series}
f = \left( 1 + R \right) \cdot \left(  a - \vb^t I \vc - \sum_{i \geq 1}\vb^T D^i \vc \right)
\end{equation}

Recall that $f$ is a homogeneous polynomial of degree $d \ge 2$. We shall now compare the homogeneous components of both sides in \eqref{eq:det-power-series}. 

Note that since the constant term of $1 +R $ is $1$, we have that if $k$ is the smallest natural number such that the degree $k$ homogeneous component of
the right hand side of \eqref{eq:det-power-series}
is non-zero, then this homogeneous component must equal the degree $k$ homogeneous component of $(a - \vb^t I \vc - \sum_{i \geq 1}\vb^T D^i \vc)$. 

If $f$ is homogeneous, then the lowest degree homogeneous component of the right hand side of \eqref{eq:det-power-series} that is non-zero must have degree equal to $d$, and must equal $f$. Moreover, since $\deg(f) \ge 2$, we get that 
\[
f = -\vb^T (D^{d-2}) \vc\, .
\]
Thus, $-\vb^T D^{d-2} \vc$ is a homogeneous ABP that computes $f$ and has the properties claimed in the lemma. 
\end{proof}

The ABP constructed in the proof of \autoref{thm:reg-det-abp} has a very special structure. Apart from the first and last layers, all the middle layers are identical and have the same transition matrix $D$. Further, $\vb, \vc$ and $D$ satisfy the equations $\vb^T D^{i} \vc = 0$ for all $0 \le i \le d-3$. Hence, for the sake of proving super-linear lower bounds on determinantal complexity, one may focus on ABPs that have this structure, although it's not clear (to us) how to utilize this additional structure to get stronger lower bounds.

\section{Algebraic Branching Programs for All Homogeneous Polynomials}
\label{sec:general}

In this section we generalize the construction given in \autoref{sec:regular} and construct ABPs of size $s \cdot \poly(d)$ for all homogeneous degree-$d$ polynomials.

\begin{theorem}
\label{thm:gen-det-abp}
Let $f \in \F[x_1, \ldots, x_n]$ be a homogeneous polynomial of degree $d$. Suppose $\dc(f) = s$. Then $\habp(f) \le O(d^5 \cdot s)$.
\end{theorem}

We begin as before by putting the determinantal representation of $f$ in a convenient normal form.

\begin{claim}
\label{cl:normal-form}
Let $f$ be a homogeneous polynomial over $\F$ of degree $d \ge 2$ and $M$ an $s \times s$ determinantal representation of $f$ over $\F$.
Write $M = M' + M_0$ where $M_0$ is the constant part of $M$, and denote $\rank(M_0)=s-r$. Then $r \ge 1$, and if $r < s$, there exists a matrix $\tilde{M}$, with $\det(\tilde{M}) = f$ such that
\begin{equation}
\label{eq:normal-form}
\tilde{M} = \begin{bmatrix}
A & B \\ C & I - D
\end{bmatrix}
\end{equation}
where $A \in \F^{r \times r}$, $B \in \F^{r \times (s-r)}$, $C \in \F^{(s-r) \times r}$, $D \in \F^{(s-r) \times (s-r)}$, $A,B,C,D$ are matrices whose entries are homogeneous linear functions, and $I$ is the $(s-r) \times (s-r)$ identity matrix.
\end{claim}

Before proving \autoref{cl:normal-form}, we note that the case $r=s$ is rather uninteresting: indeed,  if $r=s$ then $M$ is a matrix of homogeneous linear functions such that $\Det(M)$ is a homogeneous polynomial of degree $d$, which implies that $s=d$, which in turn makes the contents of \autoref{thm:gen-det-abp} trivial by applying the ABP construction of \cite{MV97} (\autoref{thm:ABP-for-det}) directly to $M$.

\begin{proof}
Since $f$ is homogeneous of degree $d \ge 1$ we must have $r \le s-1$ as otherwise $\Det(M)$ would have a non-zero constant term. The proof again follows simply by applying Gaussian elimination on the rows and columns of $M$, so that we may assume that $M_0 = \diag(0,\ldots,0,1,\ldots,1)$ where the number of $0$'s is $r$ and the number of $1$'s is $s-r$, which implies equation \eqref{eq:normal-form} by defining $A,B,C,D$ appropriately.
\end{proof}

As a corollary, we obtain the following:

\begin{corollary}
\label{cor:hom-schur-comp}
Let $f$ be a homogeneous polynomial over $\F$ of degree $d \ge 2$ and $M$ an $s \times s$ determinantal representation of $f$ over $\F$ in normal form as in \eqref{eq:normal-form}.
Expand
\[
A-B(I-D)^{-1}C = A- B\left(\sum_{i \ge 0} D^i\right)C = A - \sum_{i \ge 0} BD^i C
\]
as matrices over the ring $\F[[\vx]]$ of powers series in $\vx$  .

Then, the lowest degree non-zero homogeneous component of $\Det(A - \sum_{i \ge 0} BD^i C)$ is of degree $d$ and equals $f$.
\end{corollary}

\begin{proof}
Note that $\Det(I-D)$ is a polynomial whose constant term is $1$, i.e., $\Det(I-D) = 1 + R$ where $R$ is a constant-free polynomial. Thus implies that $I-D$ is invertible, and therefore 
by \autoref{lem:schur-complement}, 
\[
f = \Det(M) = \Det(I - D) \cdot \Det(A - B(I-D)^{-1}C) = (1+R) \cdot \Det(A - B(I-D)^{-1}C).
\]
In the power series expansion $\Det(A - B(I-D)^{-1}C) = h + Q$, where $h$ is the lowest degree non-zero homogeneous component of the powers series and $Q$ is either 0 or a power series containing only monomials of degree strictly larger than $\deg(h)$. Then
\[
f = (1+R) \cdot (h+Q) = h + P
\]
 where $P=Q+Rh+RQ$ is either 0 or contains only monomials of degree strictly larger than $\deg(h)$. Since $f$ is homogeneous of degree $d$ and $h \neq 0$, we must have $\deg(h)=d$ and $h = f$.
\end{proof}

The benefit of \autoref{cor:hom-schur-comp} is that now instead of the $s \times s$ matrix $M$ we are dealing with the $r \times r$ matrix $(A - \sum_{i \ge 0} BD^i C)$. This however comes with some costs. The first is that now we can only say that $f$ is the lowest degree homogeneous component of the determinant of this smaller matrix. The second is that this smaller matrix involves power series. The second problem is easily resolved via the following simple observation.

\begin{observation}
\label{obs:inverse-power-series}
Let $A,B,C,D$ as in \eqref{eq:normal-form} and let $d \in \N$. For a polynomial $f$, let $\Hom_d(f)$ denote its homogeneous component of degree $d$. Then
\[
\Hom_d (\Det(A-B(I-D)^{-1}C)) = \Hom_d (\Det(A-BC - BDC - BD^2C - \cdots - BD^{d-2}C)).
\]
\end{observation}

\begin{proof}
We have that $(I-D)^{-1} = I + D + D^2 + \cdots = \sum_{i=0}^\infty D^i$, so
\[
A-B(I-D)^{-1}C = A - B \left(\sum_{i=0}^\infty D^i \right) C.
\]
However, since we're interested in the degree-$d$ component of the determinant of this matrix, and $A,B,C,D$ all have homogeneous linear functions as their entries, any power of $D$ larger than $d-2$ can't contribute anything to this component (as it will only contribute monomials of degree larger than $d$), which implies we can truncate the power series at $i=d-2$ to obtain the observation.
\end{proof}

We remark that one can slightly tighten the analysis in \autoref{obs:inverse-power-series}. Since the matrix $A-BC - BDC - BD^2C - \cdots - BD^{d-2}C$ is an $r \times r$ matrix containing only constant free polynomials, when computing its degree-$d$ homogeneous component we can even truncate the power series at $BD^{d-r+2}C$. This can save a factor of $d$ in the final analysis when $r$ is very close to $d$.

The representation above already gets us very close to the final construction of the ABP.

\begin{lemma}
\label{lem:normal-form-properties}
Let $f$ be a homogeneous polynomial over $\F$ of degree $d \ge 1$ and $M$ an $s \times s$ determinantal representation of $f$ over $\F$ in normal form as in \eqref{eq:normal-form}.
Define the $r \times r$ matrix
\begin{equation}
\label{eq:smaller-form}
W = A - BC - BDC - BD^2C - \cdots - BD^{d-2}C.
\end{equation}
Then:
\begin{enumerate}
\item \label{item:degree} Every entry of $W$ is a constant free polynomial of degree at most $d$.
\item \label{item:hom-f} The smallest degree homogeneous component of $\Det(W)$ equals $f$.
\item \label{item:r} $r \le d$.
\item \label{item:abp-size} Every entry of $W$ is a polynomial computed by an ABP of size at most $O(d s)$ and of width $O(s)$.
\end{enumerate}
\end{lemma}

\begin{proof}
Items \ref{item:degree} and \ref{item:hom-f} follow from \autoref{cor:hom-schur-comp} and \autoref{obs:inverse-power-series}.

To prove item \ref{item:r}, note that by assumption every entry of $A,B,C,D$ is a homogeneous linear polynomial. Thus, $W \in \F^{r \times r}$ is a matrix of constant free polynomials, and the smallest degree homogeneous component of $\Det(W)$ is of degree at least $r$. Since it is of degree $d$ (by item \ref{item:hom-f}), $r \le d$.

Item \ref{item:abp-size} also follows easily from the definition of $W$. The $(i,j)$-th entry of $W$ is given by
\[
\ve_i W \ve_j = \ve_i A \ve_j - \ve_i BC \ve_j - \ve_i BDC \ve_j - \cdots - \ve_i BD^{d-2}C \ve_j.
\]
Each summand above is computed by an ABP of size $O(ds)$ as per \autoref{def:ABP}. Summing up those ABPs we get an ABP of size $O(d^2 s)$.

We can slightly improve the upper bound to $O(ds)$ as follows. We first compute $(I + D + \ldots + D^{d-2})$ in one shot using an ABP of size $O(ds)$ and width $O(s)$  as the top-right (block) entry of the following matrix multiplications:
\[
\left(
\begin{bmatrix}
I &I\\ 0 &D
\end{bmatrix}
\right)^{d-1}
=
\begin{bmatrix}
I &I + D + \ldots + D^{d-2}\\ 0 &D^{d-1}
\end{bmatrix}
.
\]
We then multiply by $\ve_i B$ on the left and $C\ve_j$ on the right, for a total of $O(s)$ more vertices (without increasing the width).
\end{proof}

We can now complete the proof of \autoref{thm:gen-det-abp}.

\begin{proof}[Proof of \autoref{thm:gen-det-abp}]
Let $f$ be a homogeneous polynomial over $\F$ of degree $d \ge 1$ and $M$ an $s \times s$ determinantal representation of $f$ over $\F$ in normal form as in \eqref{eq:normal-form}.
Let $W$ be as in \autoref{lem:normal-form-properties}.

We now construct the following ABP that computes $f$ using the following steps. We start by taking an ABP that computes the determinant of an $r \times r$ symbolic matrix. (where $r$ is as in \autoref{cl:normal-form}). This has size $O(r^3) = O(d^3)$ by the construction of \cite{MV97} (\autoref{thm:ABP-for-det}) and \autoref{lem:normal-form-properties}.

We now replace each variable $x_{i,j}$ by the ABP of size $O(d s)$ computing the $(i,j)$-th entry of $W$ given by \autoref{lem:normal-form-properties}. We get an ABP of total size $O(d^4 s)$ that computes $\Det(W)$.

Now, from \autoref{lem:abp-partial-homogenization}, we get that there is a homogeneous ABP of size at most $O(d^5 s)$ that computes the degree $d$ homogeneous component. 
\end{proof}

\section{$r$-Regular Determinantal Complexity}
\label{sec:r-regular}

The methods of  \autoref{sec:regular} and \autoref{sec:general} suggest that an important parameter in the study of determinantal representations is the rank of the constant part of the matrix. In this section we investigate it further and define classes of determinantal representations parametrized by this rank. These generalize the definition of a regular determinantal representation and regular determinantal complexity (\autoref{def:det-complexity}).

\begin{definition}\label{def:r-reg-det-complexity}
We define a determinantal representation $M$ of $f$ to be $r$-\emph{regular} if the constant part $M_0$ of $M$ (i.e., $M(\vzero)$) is of rank $s-r$. We denote by $\rdc_r(f)$ the minimal integer $s$ such that $f$ has a $r$-regular determinantal representation of size $s$.
\end{definition}

We emphasize that under our definition $r$ denotes \emph{co-}rank of the constant part (rather than the rank itself), as this parameter is slightly more elegant to work with. For $r = 1$, this definition is identical to the previous definition of regular representation, i.e.\ $\rdc_1(f) = \rdc(f)$.

One may again relate the rank of the constant part of determinantal representations of $f$ to natural varieties associated with $f$, as in \autoref{lemma:vzg}. Now, instead of looking at the variety defined by first-order partial derivatives, we look at partial derivatives of order up to $r$.

\begin{definition}\label{def:r-sing-locus}
Let $f \in \F[\vx]$ be a polynomial. We define $S_r(f)$ is the variety defined by
\[
S_r(f) = \left\{ \va : \frac{\partial^r f}{\partial x_{i_1}\cdots \partial x_{i_{r'}}}(\va) = 0, \text{ for every } r' \le r \text{ and for all } i_1, \ldots, i_{r'} \in \set{1,\ldots,n} \right\}.
\]
That is, $S_r(f)$ is the set of common zeros of all partial derivatives of $f$ of order at most $r$.
\end{definition}

Clearly, since partial derivatives of the determinant are either identically zero or determinants of smaller submatrices, $S_r(\Det_m)$ is the set of matrices of rank at most $m-(r+1)$.
The following fact is a generalization of \autoref{fact:vzgathen}.
\begin{fact}
\label{fact:rvzgathen}
Let $\F$ be an algebraically closed field and let $\Det_m$ be the $m \times m$ determinant polynomial. Then $S_r(\Det_m) \subseteq \F^{m\times m}$ is precisely the set of matrices of rank at
most $m - (r+1)$, and $\dim S_r(\Det_m) = m^2 - (r+1)^2$.
\end{fact}

The proof is a identical to the proof of \cite[Lemma~2.1]{vzG87} with the required changes in parameters.

\begin{proof}[Proof Sketch]
As in \cite{vzG87}, for $1\le i_1\le \ldots \le i_r \le m$, define
\[
S_{i_1, \ldots, i_r} = \{M\in \F^{m\times m}: \text{rows~} i_1,\ldots, i_r \text{~are linearly dependent on the other rows of~} M\}.
\]

We can therefore write $S_{m-r+1, \ldots, m}$, for example, as the image of the following mapping.
\[
\phi: \F^{(m-r)\times m}\times \F^{m-r}\times \cdots\times \F^{m-r}\to \F^{m\times m},
\]
\[
(M', u_1, \ldots, u_r)\mapsto 
\begin{bmatrix}
M'\\
u^T_1M'\\
\vdots\\
u^T_rM'
\end{bmatrix}.
\]
The rest of the proof follows through as in \cite[Lemma~2.1]{vzG87}.
\end{proof}

We can now conclude as before the following natural analog of \autoref{lemma:vzg}.

\begin{lemma}\label{lemma:r-vzg}
Let $f \in \F[\vx]$ be a polynomial, and let $M : \F^n \to \F^{s\times s}$ be a polynomial map
such that $f(\vx) = \Det_s(M(\vx))$. Suppose further that $\dim S_r(f) < n - (r+1)^2$. Then $\Img(M) \cap
S_r(\Det_s) = \emptyset$. Furthermore, all matrices in $\Img(M)$ have rank at least $s - r$.
\end{lemma}

\begin{proof}[Proof Sketch]
We follow the proof of \cite[Theorem 3.1]{vzG87} (see also \cite[Lemma 3.5]{KV22}). Suppose $f(\vx) = \Det_s(M(\vx))$ and suppose $A \in \Img(M) \cap S_r(\Det_s)$ so that $A = M(\vz)$ for some $\vz \in \F^n$. We claim that $\vz \in S_r(f)$, which implies that $M^{-1} (S_r(\Det_s)) \subseteq S_r(f)$ so that
\[
\dim(M^{-1} (S_r(\Det_s))) \le \dim(S_r(f)) < n-(r+1)^2,
\]
but on the other hand, since $\Img(M)$ and $S_r(\Det_s)$ aren't disjoint, by \autoref{fact:rvzgathen} and by Theorem 17.24 of \cite{HarrisAlgebraicGeometry}
\[
\dim(M^{-1} (S_r(\Det_s))) \ge n+(s^2 - (r+1)^2) - s^2 = n-(r+1)^2
\]
which is a contradiction (and the ``moreover'' part of the lemma follows from \autoref{fact:rvzgathen}).

It remains to show that $\vz \in S_r(f)$. Consider any partial derivative of $f$ of order $r' \le r$, $\frac{\partial^r f}{\partial x_{i_1} \partial x_{i_2} \cdots \partial x_{i_{r'}}}$. Since $f(\vx) = \Det(M(\vx))$, by repeated application of the chain rule and the product rule, we get that $\frac{\partial^r f}{\partial x_{i_1} \partial x_{i_2} \cdots \partial x_{i_{r'}}}$ is a sum of terms, each of which has the form 
\[
\left( \frac{\partial^t \Det_s}{\partial y_{j_1, k_1} \cdots \partial y_{j_t, k_t}} (M(\vx))\right) \cdot g
\]
where $g$ is some product of derivatives of the coordinates of $M$, for some $t \le r' \le r$ and some choices of indices $j_1, k_1, \ldots, j_t, k_t$. Since $M(\vz) = A \in S_r(\Det_s)$, this entire expression equals 0 (regardless of $g$).
\end{proof}

And again, an easy consequence of this lemma is the following observation, analogous to \autoref{obs:regular-reps}.
\begin{observation}
\label{obs:r-regular-reps}
Let $f \in \F[\vx]$ be a polynomial whose constant term is zero. Further assume that $\dim S_r(f) < n - (r+1)^2$. Then any determinantal representation of $f$ must be $r'$-regular for some $r'\leq r$.
\end{observation}

\begin{proof}
Let $f$ has a determinantal representation $M$ of size $s$. As $\dim S_r(f) < n - (r+1)^2$, using \autoref{lemma:r-vzg} we know that all matrices in $\Img(M)$ have rank at least $s - r$. In particular, $M(\vzero)$ is of rank $\geq s - r$. As $f$ does not have a constant term, $f(\vzero) = \Det(M(\vzero)) = 0$. Therefore, $s-1 \geq \rank(M(\vzero))\geq s-r$ and $M$ is an $r'$-regular determinantal representation of $f$ for some $r'\leq r$.
\end{proof}

Having defined $r$-regular determinantal representations, we remark that the construction given in \autoref{thm:gen-det-abp} of \autoref{sec:general} implies the following theorem.

\begin{theorem}
\label{thm:r-reg-det-abp}
Let $f \in \F[x_1, \ldots, x_n]$ be a homogeneous polynomial of degree $d\ge2$. Suppose $\rdc_r(f) = s$. Then $\habp(f) \le O(r^3\cdot d^2\cdot s)$. % and $\habpw(f) \le O(rs)$.
\end{theorem}

We remark that \autoref{thm:r-reg-det-abp} indeed generalizes \autoref{thm:gen-det-abp}, as degree-$d$ polynomials have only $r$-regular determinantal representations in which $r \le d$ (see item \ref{item:r} of \autoref{lem:normal-form-properties}).

\autoref{thm:r-reg-det-abp} follows directly by inspecting the proof of \autoref{thm:gen-det-abp} and keeping $r$ as a separate parameter instead of using the crude upper bound $r \le d$.

\section{Open problems}
We conclude with some open problems. 

\begin{enumerate}
\item Based on the results in this paper, a natural question to investigate further is the question of proving super linear lower bounds for ABPs for a constant degree polynomial. For such a lower bound to imply a lower bound on determinantal complexity lower bounds, we require a lower bound on the number of vertices in the ABP. Perhaps an easier first step would be to prove a super linear lower bound on the number of edges in an ABP for a constant degree polynomial.  
\item It would be very interesting to extend the tight connection between determinantal complexity and ABP size/width observed here to polynomials of large degree. Note that from \autoref{thm:reg-det-abp}, if we consider the complexity measure of ABP width, such a connection holds (independent of degree) for homogeneous polynomials that have singular loci of dimension at most $n-5$. Extending this to arbitrary polynomials in  a way that does not incur the $\poly(d)$ multiplicative blow up in size observed in \autoref{thm:gen-det-abp} would be very interesting. 
\item The notion of determinantal complexity of a polynomial can be naturally generalized in the following way: a polynomial $f \in \F[\vx]$ is said to have degree $d$ determinantal complexity (denoted by $\dc^d(f)$) at most $s$ if there is a matrix $M \in \F[\vx]^{s \times s}$ such that $\Det(M) = f$ and every entry of $M$ is a polynomial of degree at most $d$. Understanding the behavior of $\dc^d(f)$ as $d$ increases would be interesting. Besides being a natural quantity on its own, it offers a potential approach towards improving the known determinantal complexity lower bounds using the techniques in \cite{KV22}. 
\end{enumerate}

\bibliographystyle{customurlbst/alphaurlpp}
\bibliography{references}

\begin{thebibliography}{CKSV22}

\bibitem[ABV17]{ABV17}
Jarod Alper, Tristram Bogart, and Mauricio Velasco.
\newblock \href {http://dx.doi.org/10.1007/s10208-015-9300-x} {A Lower Bound
  for the Determinantal Complexity of a Hypersurface}.
\newblock {\em Found. Comput. Math.}, 17(3):829--836, 2017.

\bibitem[Ber84]{B84}
Stuart~J. Berkowitz.
\newblock \href {http://dx.doi.org/10.1016/0020-0190(84)90018-8} {On computing
  the determinant in small parallel time using a small number of processors}.
\newblock {\em Information Processing Letters}, 18(3):147 -- 150, 1984.

\bibitem[BS83]{BS83}
Walter Baur and Volker Strassen.
\newblock \href {http://dx.doi.org/10.1016/0304-3975(83)90110-X} {The
  Complexity of Partial Derivatives}.
\newblock {\em Theoretical Computer Science}, 22:317--330, 1983.

\bibitem[CCL10]{CCL10}
Jin{-}Yi Cai, Xi~Chen, and Dong Li.
\newblock \href {http://dx.doi.org/10.1007/s00037-009-0284-2} {Quadratic Lower
  Bound for Permanent Vs. Determinant in any Characteristic}.
\newblock {\em Comput. Complex.}, 19(1):37--56, 2010.

\bibitem[CKSV22]{CKSV22}
Prerona Chatterjee, Mrinal Kumar, Adrian She, and Ben~Lee Volk.
\newblock \href {http://dx.doi.org/10.1007/s00037-022-00223-8} {Quadratic Lower
  Bounds for Algebraic Branching Programs and Formulas}.
\newblock {\em Comput. Complex.}, 31(2):8, 2022.

\bibitem[Har95]{HarrisAlgebraicGeometry}
Joe Harris.
\newblock \href {http://dx.doi.org/10.1007/978-1-4757-2189-8} {{\em Algebraic
  geometry}}, volume 133 of {\em Graduate Texts in Mathematics}.
\newblock Springer-Verlag, New York, 1995.
\newblock A first course, Corrected reprint of the 1992 original.

\bibitem[Kal85]{Kal85}
Kyriakos Kalorkoti.
\newblock \href {http://dx.doi.org/10.1137/0214050} {{A Lower Bound for the
  Formula Size of Rational Functions}}.
\newblock {\em SIAM Journal of Computing}, 14(3):678--687, 1985.

\bibitem[Kum19]{K19}
Mrinal Kumar.
\newblock \href {http://dx.doi.org/10.1007/s00037-019-00186-3} {A quadratic
  lower bound for homogeneous algebraic branching programs}.
\newblock {\em Computational Complexity}, 28(3):409--435, 2019.

\bibitem[KV22]{KV22}
Mrinal Kumar and Ben~Lee Volk.
\newblock \href {http://dx.doi.org/10.1007/s00037-022-00228-3} {A Lower Bound
  on Determinantal Complexity}.
\newblock {\em Comput. Complex.}, 31(2):12, 2022.

\bibitem[KW93]{KW93}
Mauricio Karchmer and Avi Wigderson.
\newblock \href {http://dx.doi.org/10.1109/SCT.1993.336536} {On Span Programs}.
\newblock In {\em \CCC{1993}}, pages 102--111. {IEEE} Computer Society, 1993.

\bibitem[Lok09]{Lokam09}
Satyanarayana~V. Lokam.
\newblock \href {http://dx.doi.org/10.1561/0400000011} {Complexity Lower Bounds
  using Linear Algebra}.
\newblock {\em Found. Trends Theor. Comput. Sci.}, 4(1-2):1--155, 2009.

\bibitem[LR17]{LR17}
J.M. Landsberg and Nicolas Ressayre.
\newblock \href {http://dx.doi.org/10.1016/j.difgeo.2017.03.017} {Permanent v.
  determinant: An exponential lower bound assuming symmetry and a potential
  path towards Valiant's conjecture}.
\newblock {\em Differential Geometry and its Applications}, 55:146--166, 2017.

\bibitem[MR04]{MR04}
Thierry Mignon and Nicolas Ressayre.
\newblock \href {http://dx.doi.org/10.1155/S1073792804142566} {{A quadratic
  bound for the determinant and permanent problem.}}
\newblock {\em International Mathematics Research Notes}, 2004(79):4241--4253,
  2004.
\newblock Available on
  \href{http://citeseerx.ist.psu.edu/viewdoc/summary?doi=10.1.1.106.4910}{\tt
  citeseer:10.1.1.106.4910}.

\bibitem[MV97]{MV97}
Meena Mahajan and V.~Vinay.
\newblock \href {https://dl.acm.org/doi/10.5555/314161.314429} {A Combinatorial
  Algorithm for the Determinant}.
\newblock In {\em \SODA{1997}}, pages 730--738, 1997.
\newblock Available on
  \href{http://citeseerx.ist.psu.edu/viewdoc/summary?doi=10.1.1.31.1673}{\tt
  citeseer:10.1.1.31.1673}.

\bibitem[Nec66]{Nec66}
Eduard~Ivanovich Nechiporuk.
\newblock \href {http://mi.mathnet.ru/dan32449} {On a Boolean function}.
\newblock {\em Dokl. Akad. Nauk SSSR}, 169:765--766, 1966.

\bibitem[Raz10]{raz-elusive}
Ran Raz.
\newblock \href {http://dx.doi.org/10.4086/toc.2010.v006a007} {Elusive
  Functions and Lower Bounds for Arithmetic Circuits}.
\newblock {\em Theory of Computing}, 6(7):135--177, 2010.

\bibitem[Sap15]{S15}
Ramprasad Saptharishi.
\newblock \href {https://github.com/dasarpmar/lowerbounds-survey/releases/} {A
  survey of lower bounds in arithmetic circuit complexity}.
\newblock Github survey, 2015.

\bibitem[SS97]{ShoupS1997}
Victor Shoup and Roman Smolensky.
\newblock \href {http://dx.doi.org/10.1007/BF01270384} {Lower Bounds for
  Polynomial Evaluation and Interpolation Problems}.
\newblock {\em Comput. Complex.}, 6(4):301--311, 1997.

\bibitem[Str73]{Strassen73}
Volker Strassen.
\newblock \href {http://dx.doi.org/10.1007/BF01436566} {Die
  Berechnungskomplexit\"{a}t Von Elementarsymmetrischen Funktionen Und Von
  Interpolationskoeffizienten}.
\newblock {\em Numerische Mathematik}, 20(3):238--251, June 1973.

\bibitem[Val75]{Valiant75}
Leslie~G. Valiant.
\newblock \href {http://dx.doi.org/10.1145/800116.803752} {On Non-linear Lower
  Bounds in Computational Complexity}.
\newblock In {\em Proceedings of the 7th Annual {ACM} Symposium on Theory of
  Computing, May 5-7, 1975, Albuquerque, New Mexico, {USA}}, pages 45--53.
  {ACM}, 1975.

\bibitem[Val77]{Valiant77}
Leslie~G. Valiant.
\newblock \href {http://dx.doi.org/10.1007/3-540-08353-7\_135} {Graph-Theoretic
  Arguments in Low-Level Complexity}.
\newblock In {\em \MFCS{1977}}, volume~53 of {\em Lecture Notes in Computer
  Science}, pages 162--176. Springer, 1977.

\bibitem[Val79]{Val79}
Leslie~G. Valiant.
\newblock \href {http://dx.doi.org/10.1145/800135.804419} {{Completeness
  Classes in Algebra}}.
\newblock In {\em \STOC{1979}}, pages 249--261, 1979.

\bibitem[vzG87]{vzG87}
Joachim von~zur Gathen.
\newblock \href {https://core.ac.uk/download/pdf/82095887.pdf} {Permanent and
  determinant}.
\newblock {\em Linear Algebra and its Applications}, 96:87--100, 1987.

\end{thebibliography}

\end{document}